\documentclass[osajnl,twocolumn,showpacs,10pt]{revtex4-1}
\usepackage{amsmath,amssymb,graphicx}
\begin{document}

\title{Feedback-enhanced algorithm for aberration correction of holographic atom traps}
\author{Graham D. Bruce, Matthew Y. H. Johnson, Edward Cormack, David Richards, James Mayoh and Donatella Cassettari}\email{Corresponding author: dc43@st-andrews.ac.uk}
\affiliation{Scottish Universities Physics Alliance, School of Physics and Astronomy, University of St Andrews, St Andrews, Fife KY16 9SS, UK }

\begin{abstract}
We show that a phase-only spatial light modulator can be used to generate non-trivial light distributions suitable for trapping ultracold atoms, when the hologram calculation is included within a simple and robust feedback loop that corrects for imperfect device response and optical aberrations. This correction reduces the discrepancy between target and experimental light distribution to the level of a few percent (RMS error). We prove the generality of this algorithm by applying it to a variety of target light distributions of relevance for cold atomic physics.
\end{abstract}


\maketitle



A recent area of interest in the field of cold atomic physics is the development of non-trivial spatially- and temporally-varying optical trapping geometries, with interesting examples already demonstrated using techniques including acousto-optic deflection \cite{Houston_08,Henderson_09,Zimmermann_11,
Trypogeorgos_13}, amplitude- \cite{Muldoon_12,Lee_14} and phase-modulation \cite{McGloin_03,Bergamini_04,Boyer_06,
Franke-Arnold_07,Bruce_11ring,Bruce_11power,
Gaunt_12,Lee_14} of trapping light.  Optical traps generally offer increased trap complexity at small length-scales, but at the disadvantage of increased likelihood of small-scale potential roughness \cite{Pasienski_08}.  Any local roughness in the intensity of the light pattern creates a varying energy landscape, which could cause heating or fragmentation of the atom cloud \cite{fortagh_02}. 


Fourier-engineered optical traps (those based on phase-only spatial modulation of the light to tailor the intensity in the Fourier plane of an optical system) have predominantly taken the form of arrays of discrete traps \cite{Bergamini_04,Boyer_06} or Laguerre-Gauss beams \cite{Franke-Arnold_07}.  Recently, a new calculation method for phase-only holograms of arbitrary complexity directly addressed the issue of roughness. This algorithm, the Mixed-Region Amplitude Freedom (MRAF) \cite{Pasienski_08} variant of the Gerchberg--Saxton iterative Fourier transform algorithm \cite{Gerchberg_72}, calculates smooth and accurate light patterns for use as optical atom traps.  However, other than in special cases \cite{Bruce_11power}, the output of this algorithm, when applied to real devices, does not give high-quality optical traps and this output must be further adjusted \cite{Bruce_11ring,Gaunt_12}. In this letter, we present a simple, robust and generally-applicable algorithm to improve the accuracy of optical traps generated by phase-only spatial light modulators (SLMs).

\begin{figure}[hbp]
\centerline{\includegraphics[width=.8\columnwidth]{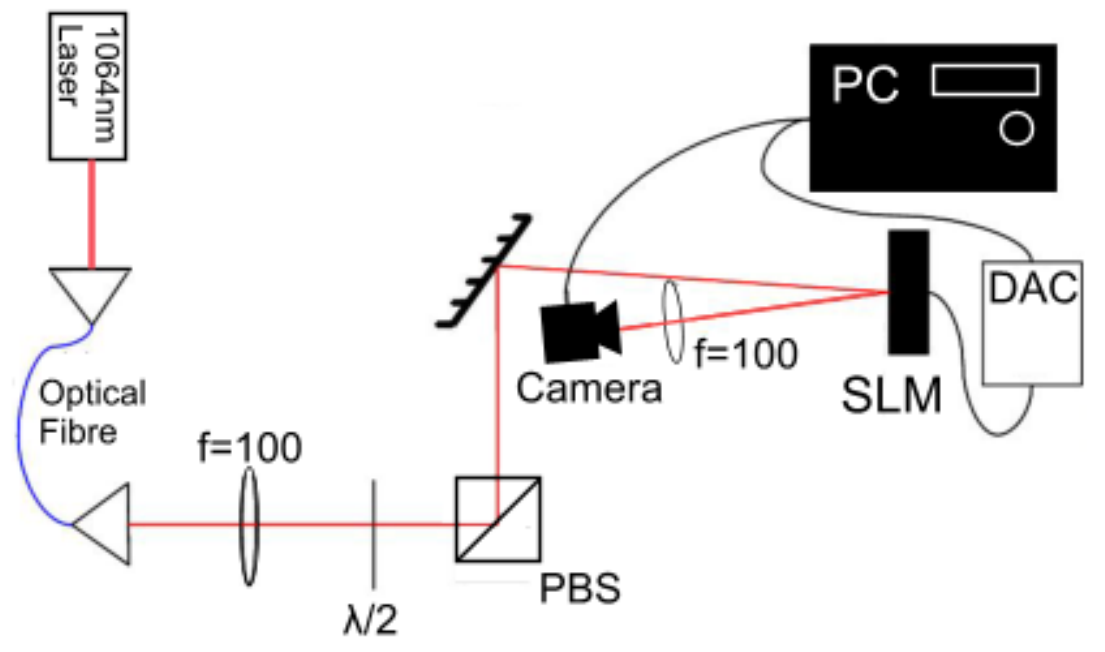}}
\caption{Experimental apparatus for feedback--enhancement of holographically generated light patterns.  A collimated Gaussian-profile laser beam illuminates a phase-only nematic liquid crystal SLM, which imparts the calculated phase modulation on the beam.  The modulated light is then collected by a lens and the intensity in the focal plane of the lens (i.e. the output plane) is recorded by a CCD camera.}
\label{fig:apparatus}
\end{figure}


The phase modulation required to produce the optical traps presented in this work is initially calculated using the MRAF algorithm.  For a given target pattern $T_{1}$ in the Fourier (i.e. output) plane, this algorithm iteratively optimises a proposed phase-only hologram by emphasising accuracy of the electric-field amplitude within a subset of the output plane (known as the Signal Region).  This target amplitude should contain the pattern of interest plus a surrounding area with zero amplitude. The amplitude is unconstrained in the remainder of the output plane (the Noise Region). The zero-amplitude region between the target pattern and the unconstrained amplitude ensure that atoms trapped in the Signal Region cannot tunnel into whatever intensity distribution is generated in the Noise Region. Calculations of trap quality are performed considering only non-zero pixels within the Signal Region (a subset known as the Measure Region, which contains the target).  Upon stagnation of the optimisation routine, the algorithm returns a phase pattern $\phi_{1}$ and a predicted amplitude $P_{1}$ which closely resembles $T_{1}$. 



Details of our apparatus, which has been introduced in \cite{Bruce_11power,Bruce_11ring}, are shown in Fig. 1.  A free-running diode laser emitting at 1060 nm injects a single-mode optical fibre to give a Gaussian beam profile.  This beam profile is expanded to a $1/e^2$ waist of $6.04$ mm, linearly polarized by a polarising beam splitter and steered onto the SLM with a narrow ($6.5^{\circ}$) incidence angle.  The reflected and phase-modulated light is focussed by a $f=100$ mm achromatic doublet and the intensity in the focal plane is collected by a CCD camera.  Both the SLM (via a digital to analogue converter) and the camera are interfaced to the same control computer. 


\begin{figure}[htbp]
\centerline{\includegraphics[width=.9\columnwidth]{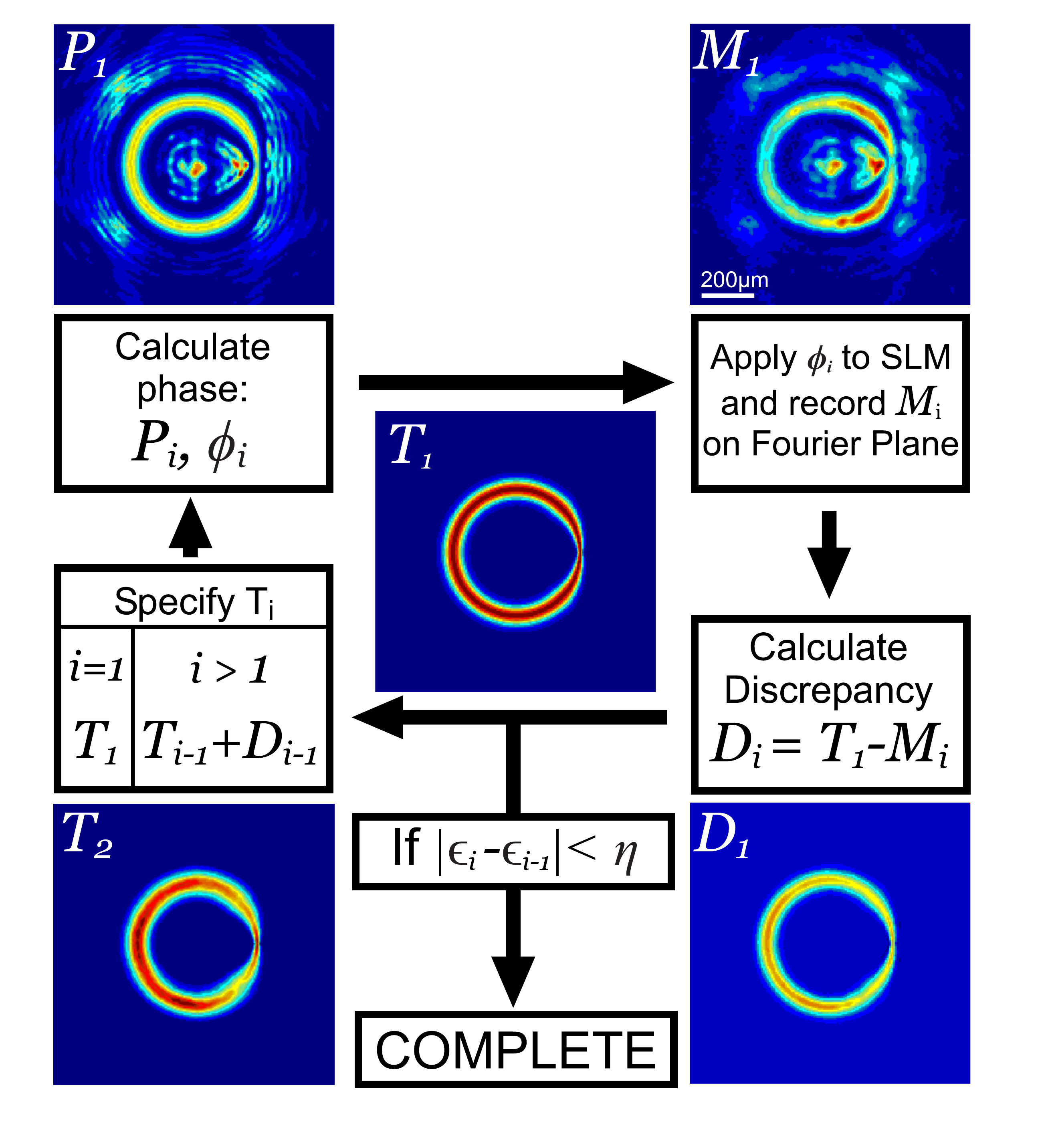}}
\caption{Block diagram outlining the basic principle of the feedback algorithm, showcasing the example of a ring trap with a restriction.}



\label{fig:flow}
\end{figure}

The phase $\phi_{1}$ is applied to the SLM and an image of the resultant Fourier--plane intensity $M_{1}$ recorded.  This output typically significantly deviates from $T_{1}$, due to aberrations in the optical system and imperfect device response.  To solve the problem of this deviation, we have designed a simple iterative feedback algorithm to steer the light pattern within the Signal Region towards the initial target $T_{1}$, as shown in Fig. \ref{fig:flow}. The error signal for  this feedback loop, the discrepancy $D_{i}$, is quantified within the Measure Region as $D_{i}=\tilde{T}_{1}-\tilde{M}_{i}$, where the tilde signifies normalisation. We normalise by the mean value of all pixels in the output brighter than $50\%$ of the maximum target value. We choose this normalization as it is resistant to both low-level background noise and particularly bright, single-pixel, noise.  A corrected target pattern $\tilde{T}_{i+1}=\tilde{T}_{i}+D_{i}$ is then generated which compensates for the discrepancies, where $i$ denotes the iteration number of the loop. This new target serves as the input for another iteration of MRAF, and generates a new phase $\phi_{i+1}$. The whole process is repeated until the output pattern reaches a stagnation point, defined to be after three subsequent iterations of feedback which improve the root mean square (RMS) error $\epsilon_{i}$ by less than a set tolerance value $\eta$, i.e. when $\left|\epsilon_{i}-\epsilon_{i-1}\right|<\eta$ for three subsequent values of $i$. For the examples below we have set $\eta=0.01\%$.

In practice we have found that increased accuracy can be achieved by introducing a gain parameter $\alpha$ to the feedback loop which is scheduled such that $\alpha>0.5$ for initial iterations to correct for large discrepancies and can be decreased as the improvement between iterations stagnates in order to impart more finely-tuned corrections.  Thus, 
 $\tilde{T}_{i+1}$ becomes $\tilde{T}_{i+1}=\tilde{T}_{i}+\alpha D_{i}$ where $\alpha$ is empirically optimized and typically ranges from $0.3$ to $0.6$ for later iterations depending on the pattern.



\begin{figure}[htbp]
\centerline{\includegraphics[width=.97\columnwidth]{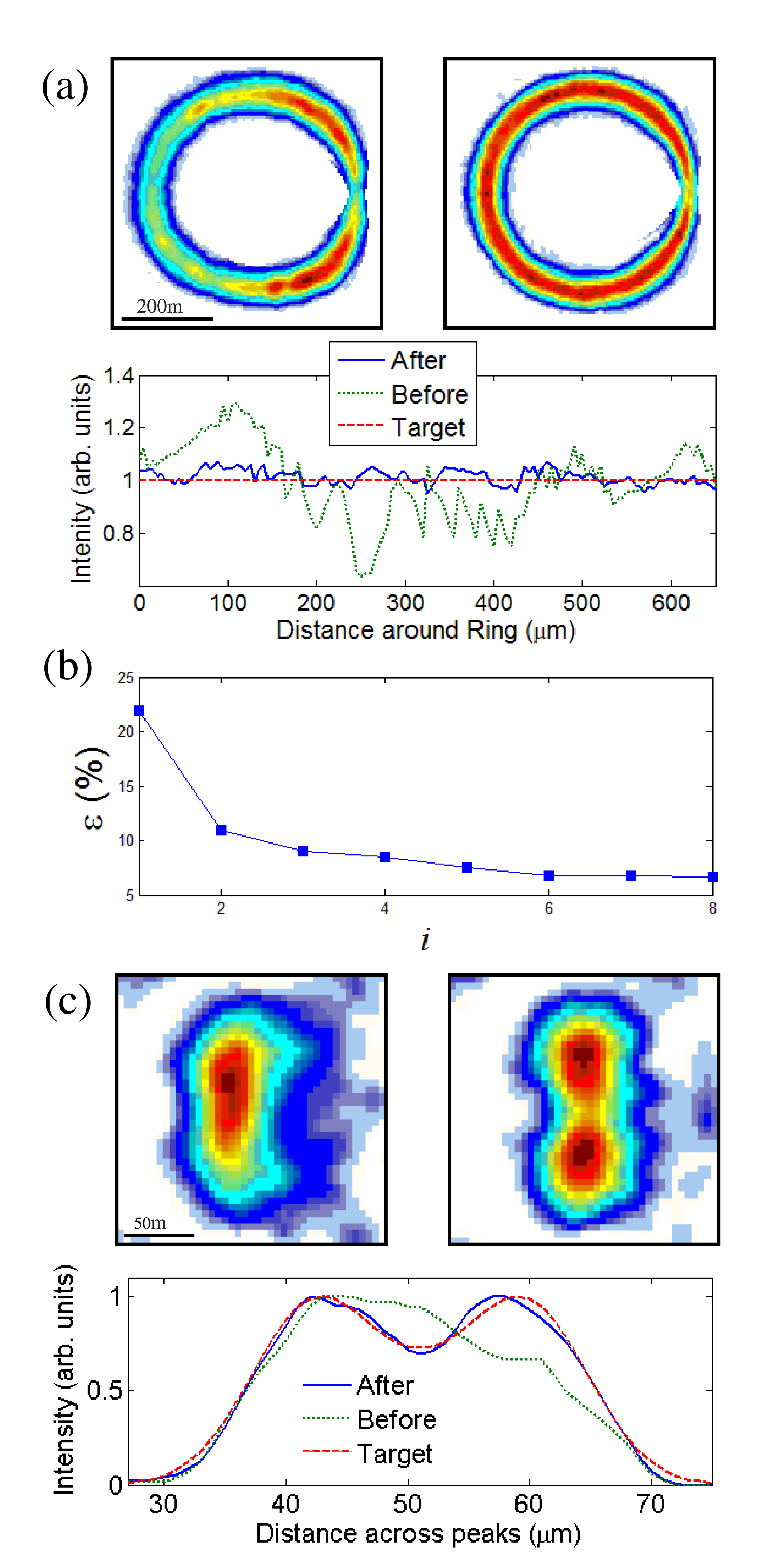}}
\caption{Continuous patterns optimised by feedback. 
(a) A ring pattern with a Gaussian radial distribution and a restriction (as in Fig. \ref{fig:flow}), showing the measured Signal Region intensity using the initial MRAF-calculated phase profile (left) and after 8 iterations of feedback optimisation (right), which improve $\epsilon^{\left(10\%\right)}$ from $9.8\%$ to $2.1\%$. (Below) The intensity around the circumference of the ring showing the target pattern (red), and the measured profile before (green) and after (blue) feedback.  (b) RMS error progression for the ring pattern with restriction, recorded at each step of the feedback algorithm.  (c) A  Gaussian double-well pattern, showing Signal-Region intensity before (left) and after (right) feedback. (Below) A cut across the centre of the two wells, showing normalised pixel intensity for the initial, final and target patterns.  For this pattern, $\epsilon^{\left(10\%\right)}$ improved from $5.5\%$ to $0.7\%$ within $10$ iterations.  
}
\label{fig:cont}
\end{figure}

\begin{figure}[htbp]
\centerline{\includegraphics[width=1\columnwidth]{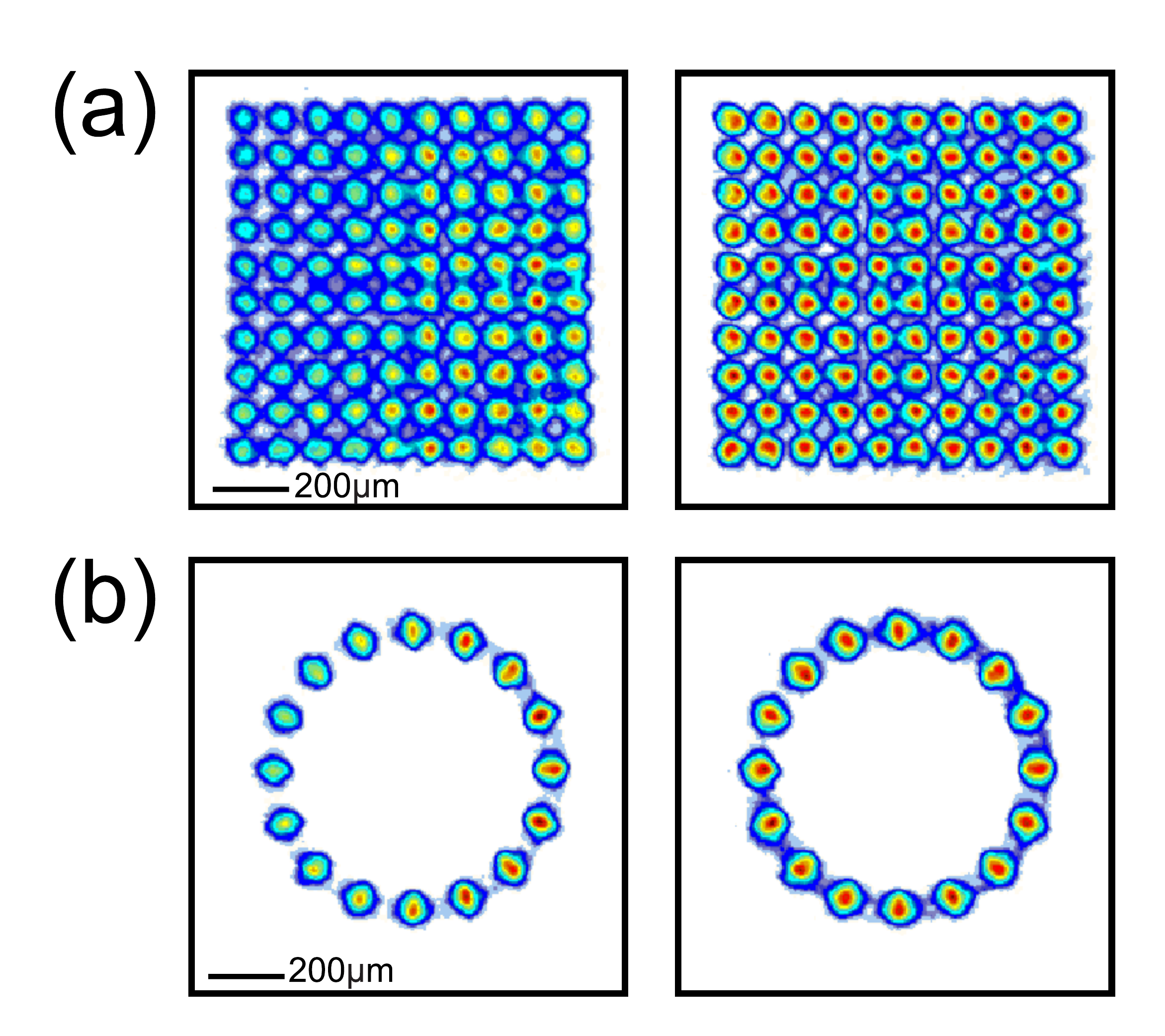}}
\caption{Discrete patterns. Signal Region intensity of (a) a $10 \times 10$ square lattice and (b) a 16-spot ring lattice before (left) and after (right) feedback.  Errors in the trapping minimum decrease from $3.6\%$ to $2.3\%$ for the square lattice and from $2.5\%$ to $1.1\%$ for the ring lattice.
}
\label{fig:disc}
\end{figure}

We test the feedback algorithm on a variety of patterns of interest, including a ring with a restriction, a Gaussian double-well and various arrays of discrete spots.  After few iterations of the feedback loop, the measured light profiles shown in Fig. \ref{fig:cont} (continuous light patterns) and Fig. \ref{fig:disc} (discrete spot patterns) show increased accuracy and more closely resemble their target patterns. A summary of the improvements to these optical traps due to the feedback process, along with the number of iterations required ($\iota$) and the light-usage efficiency ($\Gamma$), can be found in Table~\ref{tab:res}.

\begin{table}[ht]
\caption{Summary of improvements, including error, iterations ($\iota$) and Efficiency ($\Gamma$)}
  \begin{center}
    \begin{tabular}{cccp{3pt}cccc}
    \hline
    ~ & \multicolumn{2}{c}{$\epsilon~\left[\%\right]$ } & & \multicolumn{2}{c}{$\epsilon^{\left(10\%\right)}~\left[\%\right]$} & ~ & ~ \\
   \cline{2-3} \cline{5-6} 
    Pattern &  Before &  After & &  Before &  After  & $\iota$ & $\Gamma$ $\left[\%\right]$\footnote{Efficiency is the percentage of light incident on the SLM that transforms into the trapping potential.} \\
    \hline
    Indented Ring & 21.9 & 6.7 && 9.8 & 2.1 & 8 & 21.5\\
    Double Well & 19.0 & 5.5 && 5.5 & 0.7 & 10 & 20.0 \\[3pt]
    Square Lattice & 22.0 & 15.3 && 3.6 & 2.3 & 10 & 14.8 \\
    Ring Lattice & 13.7 & 10.0 && 2.5 & 1.1 & 10 & 17.5 \\
    \hline
    \end{tabular}
  \end{center}
  \label{tab:res}
\end{table}

As an evaluation metric we use the RMS error,
\begin{equation}
\epsilon_{i}=\sqrt{\frac{1}{N_{M\!R}}\sum_{M\!R}\left(\tilde{M}_{i}-\tilde{T}_{i}\right)^2},
\end{equation}
where $N_{M\!R}$ denotes the number of pixels in the measure region.  In real optical traps, atoms with thermal energy greater than $10\%$ of the trap depth will quickly evaporate from the trap \cite{ohara_01}. Following this, we also include the RMS error $\epsilon^{\left(10\%\right)}$ of only those pixels within $10\%$ of the brightest pixel - the trapping minimum - as another figure of merit. Finally, light-usage efficiencies $\Gamma$ have been calculated for each of the final output patterns. Around $50\%$ of the light incident on the SLM in our experimental set-up is diffracted into the 1st order and goes on to make up the hologram (Signal and Noise regions). To find the overall trap efficiency, we multiply the efficiency of the hologram (the percentage of 1st order light in the Signal Region) with the $50\%$ SLM efficiency.

The measured Signal Region intensity for the indented ring before and after feedback is shown in Fig. \ref{fig:cont}(a). The main deviation in the initial output is a left-right intensity gradient due to the SLM reduced diffraction efficiency at larger deflection angles. This initial pattern has $\epsilon = 21.9\%$, and a $\epsilon^{\left(10\%\right)} = 9.8\%$ .  The feedback process corrects for this intensity gradient and also corrects the width of the ring, improving $\epsilon$ to $6.7\%$ and $\epsilon^{\left(10\%\right)}$ to $2.1\%$. Along the ring of highest intensity, large fluctuations have been suppressed by the feedback process, as shown in the plot in Fig. \ref{fig:cont}(a).  A stagnation point is reached after $8$ iterations of the feedback loop, as plotted in Fig. \ref{fig:cont}(b). In this example the value of $\alpha$ changes from $0.6$ to $0.3$ after the third iteration. This evolution of the RMS error is typical for most of our feedback optimisations, showing a major improvement in the first iteration, followed by smaller improvements converging to a more accurate pattern. Most patterns are optimised in fewer than $10$ iterations of feedback.  

The final value of $\epsilon^{\left(10\%\right)}$ is sufficiently low for experiments of interest in this atom ring-trap.  If a ring with $80\mu$m radius and $12\mu$m width is generated with 3.5mW of laser power, the trap depth $U_{0}$ is $69$nK.  If a Bose--Einstein condensate of $10^5$ $^{87}$Rb atoms is trapped in this ring at zero magnetic field, the chemical potential is sufficiently low ($\mu=6.9$nK) that the atoms will be confined.  In particular, this trap can be used for studies of superfluid effects; fluctuations of $2.1\%$ in the trap depth are smaller than $\mu/5$, which is sufficiently low that the superfluidity of the gas persists \cite{Ryu_2007}.

 
A pattern which particularly demonstrates the robustness of the algorithm is the Gaussian double well shown in Fig. \ref{fig:cont}$\left(c\right)$, which has many uses for investigating fundamental quantum mechanics \cite{Milburn_97}. Before feedback the initial output is aberrated to the extent it resembles a single-well potential with an $\epsilon$ of $19.0\%$. After $10$ iterations (during which $\alpha$ is fixed at $0.6$) we obtain an $\epsilon$ of $5.5\%$ and the two wells are clearly distinguishable, $\epsilon^{\left(10\%\right)}$ having improved from $5.5\%$ to $0.7\%$. The graph shows the improvement between the initial and final outputs through the centre of the pattern. 

In Fig.~\ref{fig:disc} the target patterns are different arrangements of simple Gaussian spots of the same intensity. The simple square array is analogous to an optical lattice with the underlying spatially-varying potential removed, while the ring lattice is an experimental geometry which is interesting for quantum simulation \cite{Olomos_11,Kaminishi_11}, but is an optical lattice which cannot be created using more conventional methods such as standing waves. Our feedback algorithm corrects the size and position of aberrated spots within $10$ iterations. We measure a decrease in $\epsilon$ from $22.0\%$ to $15.3\%$ for the square lattice, and from $13.7\%$ to $10.0\%$  for the ring lattice, whilst $\epsilon^{\left(10\%\right)}$ is reduced from $3.6\%$ to $2.3\%$ for the square lattice, and from $2.5\%$ to $1.1\%$  for the ring lattice. 




In summary, the feedback algorithm is sufficienty robust to correct for large aberrations in the experimentally generated optical traps within a small number of iterations, bringing optical trap discrepancies to the percent level. Smaller-scale errors such as optical vortices can cause discrepancies in the output plane which are not compensated by the feedback algorithms.  However, these can be overcome by combining the feedback algorithm with a hologram--calculation algorithm which can directly penalise optical vortex formation, such as the recently proposed Conjugate Gradient Optimisation algorithm \cite{Harte_14}. Indeed, we have already performed initial tests of the compatibility of these two methods, with promising results.

Improvements may be made to the feedback algorithm upon its integration into a cold atoms experiment.  The aforementioned sensitivity of cold atoms to any trapping potential roughness means that we envisage continuing to use the feedback loop by taking \emph{in-situ} images of the trapped atoms \cite{Muldoon_12,Andrews_96} rather than directly imaging the light profile. Phase modulation provides precise control of the intensity in the trapping plane, but the behaviour of the intensity out of the focal plane is unpredictable, and does not necessarily diverge quickly enough to provide confinement in all directions. To provide stable three dimensional confinement when we integrate the holographic optical traps into a cold atoms experiment, a light sheet can be applied orthogonal to the trap beam to provide tight confinement. Furthermore, recently there has been significant progress developing high numerical aperture microscope objectives for cold-atoms \cite{Alt_02,Bakr_09,Sherson_10,Zimmermann_11} which could be combined with our approach to produce more finely detailed traps.

\acknowledgments{The authors wish to acknowledge helpful conversations and experimental assistance from S. L. Bromley, T. Harte, G. Smirne and L. Torralbo-Campo, and funding from EPSRC UK and the Leverhulme Trust Research Program Grant RPG-2013-074.}


\begin{thebibliography}{10}

\bibitem{Houston_08}
N.~Houston, E.~Riis, and A.~S. Arnold. Reproducible dynamic dark ring lattices for ultracold atoms J. Phys. B {\bf 41}, 211001 (2008).

\bibitem{Henderson_09}
K.~Henderson, C.~Ryu, C.~MacCormick, and M.~G. Boshier.
Experimental demonstration of painting arbitrary and dynamic potentials for Bose--Einstein condensates.
 New J. Phys. {\bf 11}, 043030 (2009).

\bibitem{Zimmermann_11}
B.~Zimmermann, T.~M\"uller, J.~Meineke, T.~Esslinger, and H.~Moritz. High-resolution imaging of ultracold fermions in microscopically tailored optical potentials. New J. Phys. {\bf 13}, 043007 (2011).

\bibitem{Trypogeorgos_13}
D.~Trypogeorgos, T.~Harte, A.~Bonnin, and C.~Foot.
Precise shaping of laser light by an acousto-optic deflector. Opt. Express {\bf 21}, 24837–24846 (2013).

\bibitem{Muldoon_12}
C.~ Muldoon, L.~Brandt, J.~Dong, D.~Stuart, E.~Brainis,
M.~Himsworth, and A.~Kuhn. Control and manipulation of cold atoms in optical tweezers. New J. Phys. {\bf 14},073051 (2012).

\bibitem{Lee_14}
J.~G. Lee and W.~T. {Hill III}.
\newblock Spatial shaping for generating arbitrary optical dipoles traps for ultracold degenerate gases. (2014). \url{http://arxiv.org/abs/1406.4084}.

\bibitem{McGloin_03}
D.~McGloin, G.~Spalding, H.~Melville, W.~Sibbett, and K.~Dholakia.
\newblock Applications of spatial light modulators in atom optics. Opt. Express {\bf 11}, 158--166 (2003).

\bibitem{Bergamini_04}
S.~Bergamini, B. Darqui\'{e}, M.~Jones, L.~Jacubowiez, A.~Browaeys, and P.~Grangier. Holographic generation of microtrap arrays for single atoms by use of a programmable phase modulator.
J. Opt. Soc. Am. B {\bf 21}, 1889--1894 (2004).

\bibitem{Boyer_06}
V.~Boyer, R.~M. Godun, G.~Smirne, D.~Cassettari, C.~M. Chandrashekar, A.~B.
  Deb, Z.~J. Laczik, and C.~J. Foot.
Dynamic manipulation of Bose-Einstein condensates with a spatial light modulator.
Phys. Rev. A {\bf 73}, 031402 (2006).

\bibitem{Franke-Arnold_07}
S. Franke-Arnold, J. Leach, M. J. Padgett, V. E. Lembessis, D. Ellinas, A. J. Wright, J. M. Girkin, P. Öhberg, and A. S. Arnold. Optical ferris wheel for ultracold atoms.
Opt. Express {\bf 15}, 8619--8625 (2007).

\bibitem{Bruce_11ring}
G.~D.~Bruce, J.~Mayoh, G.~Smirne, L.~Torralbo-Campo, and D.~Cassettari. A smooth, holographically generated ring trap for the investigation of superfluidity in ultracold atoms. Physica Scripta {\bf T143}, 014008 (2011).

\bibitem{Bruce_11power}
G.~D.~Bruce, S.~L.~Bromley, G.~Smirne, L.~Torralbo-Campo and  D.~ Cassettari. Holographic power-law traps for the efficient production of Bose-Einstein condensates. Phys. Rev. A {\bf 84}, 053410 (2011).

\bibitem{Gaunt_12}
A.~L.~Gaunt and Z.~Hadzibabic. Robust digital holography for ultracold atom trapping. Sci. Rep. {\bf 2} 721 (2012).




\bibitem{Pasienski_08}
M.~Pasienski and B.~DeMarco. A high-accuracy algorithm for designing arbitrary holographic atom traps. Opt. Express {\bf 16}, 2176 (2008).

\bibitem{fortagh_02}
J.~Fort\'agh, H.~Ott, S.~Kraft, A.~G\"unther, and C.~Zimmermann. Surface effects in magnetic microtraps.  Phys. Rev. A {\bf 66}, 041604 (2002).

\bibitem{Gerchberg_72}
R.~W. Gerchberg and W.~O. Saxton. A practical algorithm for the determination of the phase from image and difraction plane pictures. {\bf 35}, 237--246 (1972).

\bibitem{Milburn_97}
G.~J. Milburn, J.~Corney, E.~M. Wright, and D.~F. Walls. Quantum dynamics of an atomic bose-einstein condensate in a double-well potential. Phys. Rev. A {\bf 55}, 4318--4324 (1997).

\bibitem{ohara_01}
K. M. O'Hara, M. E. Gehm, S. R.Granade and J. E. Thomas. Scaling laws for evaporative cooling in time-dependent optical traps. Phys. Rev. A {\bf 64}, 051403 (2001).

\bibitem{Ryu_2007}
C. Ryu, M. F. Andersen, P. Clad\'e, V. Natarajan, K. Helmerson and W. D. Phillips. Observation of persistent flow of a Bose--Einstein condensate in a toroidal trap. Phys. Rev. Lett. {\bf 99}, 260401 (2007).

\bibitem{Olomos_11}
B. Olmos and I. Lesanovsky. Rydberg rings. Phys. Chem. Chem. Phys. {\bf 13}, 4208--4219 (2011).

\bibitem{Kaminishi_11}
E. Kaminishi, R. Kanamoto, J. Sato, and T. Deguchi. Exact yrast spectra of cold atoms on a ring. Phys. Rev. A {\bf 83}, 031601 (2011).

\bibitem{Harte_14}
T.~Harte, G.~D. Bruce, J.~Keeling, and D.~Cassettari. A conjugate gradient minimisation approach to generating holographic traps for ultracold atoms. (2014).
\newblock \url{http://arxiv.org/abs/1408.0188}.

\bibitem{Andrews_96}
M.~R. Andrews, M.-O. Mewes, N.~J. van Druten, D.~S. Durfee, D.~M. Kurn, and W.~Ketterle. Direct, Nondestructive Observation of a Bose Condensate. Science {\bf 273}, 84--87 (1996).

\bibitem{Alt_02}
W.~Alt. An objective lens for efficient fluorescence detection of single atoms. Optik {\bf 113}, 142 -- 144 (2002).

\bibitem{Bakr_09}
W.~S. Bakr, J.~I. Gillen, A.~Peng, S.~F\"{o}lling, and M.~Greiner. A quantum gas microscope for detecting single atoms in a Hubbard-regime optical lattice. Nature {\bf 462}, 74--77 (2009).

\bibitem{Sherson_10}
J.~F. Sherson, C.~Weitenberg, M.~Endres, M.~Cheneau, I.~Bloch, and S.~Kuhr. Single-atom-resolved fluorescence imaging of an atomic Mott insulator.  Nature {\bf 467}, 68--72 (2010).

\end{thebibliography}

%
%

\end{document}